# Robustness of spin polarization against temperature in multilayer structure: Triple quantum well

S. Ullah,[1, a)] F. C. D. Moraes,[1] G. M. Gusev,[1] A. K. Bakarov,[2] and F. G. G. Hernandez[1]

[1)]*Instituto de Física, Universidade de São Paulo, Caixa Postal 66318 - CEP 05315-970, São Paulo, SP, Brazil*

[2)]*Institute of Semiconductor Physics and Novosibirsk State University, Novosibirsk 630090, Russia*

(Dated: 25 April 2018)

We address the temperature influence on the precessional motion of electron spins under transverse magnetic field, studied in a GaAs/AlGaAs triple quantum wells, using pump-probe Kerr rotation. In the presence of an applied in-plane magnetic field the TRKR measurements show the robustness of carrier's spin polarization against temperature which can be easily traced in an extended range up to 250 K. By tuning the pump-probe wavelength to the exciton bound to a neutral donor transition, we observed a remarkably long-lasting spin coherence (with dephasing time $T_2^* > 14$ ns) limited by the spin hopping process and exchange interaction between the donor sites as well as the ensemble spread of $g$-factor. The temperature dependent spin dephasing time revealed a double linear dependence due to the different relaxation mechanisms active at respective temperature ranges. We observed that the increase of sample temperature from 5 K to 250 K, leads to a strong $T_2^*$ reduction by almost 98%/97% for the excitation wavelengths of 823/821 nm. Furthermore, we noticed that the temperature increase not only causes the reduction of spin lifetime but can also lead to the variation of electron $g$-factor. Additionally, the spin dynamics was studied through the dependencies on the applied magnetic field and optical pump power.

## I. INTRODUCTION

Recently, the spin dynamics of carrier's and related physics in the low-dimensional structure have attracted considerable attention from both viewpoints of physics, and it's promising applications in spintronics devices.[1–4] Long-lasting spin coherence, persisting up to about room temperature, is one of the key requirement for successful implementation of novel spintronics devices. For that reason, advanced and new material structures exhibiting large spin polarization are highly desirable. A number of efforts have been put forth to enhance the spin lifetime, for example, by using different dimensionally semiconductor nanostructures, like QWs[5,6], quantum dots (QDs)[7] and layered structures[8] of various material systems based on III-V (e.g., GaAs, GaN, (In,Ga)As) and II-VI (e.g., CdTe, ZnSe, (Zn,Cd)Se) semiconductors.

Based on those efforts, two approaches for the tailoring of carriers spin polarization have emerged. The first relies on the doping of the material, which guarantees the long spin coherence time while the second is based on the tailoring through spin-orbit (SO) field. One of those attempts made by Awschalom group in the bulk[9] and II-VI QW[10] samples, with doping level close to metal-insulator transition (MIT), was the observation of an extraordinary long coherence time. Those findings, on the one hand, revealed that the long-lived spin coherence is restrained to a doping level in the vicinity of MIT.[11–13] On the other hand, it animated the expectation that the electron spin can be finally realized as a basis for quantum computation. For the device applications, it is highly desirable that the generation and detection of such spin polarization could be carried out at room temperature and low magnetic field.

While providing a control knob for handling the carrier spins, the spin-orbit coupling (SOC) also leads to an efficient spin relaxation through the Dyakonov-Perel (DP) mechanism.[14] In this mechanism, the random walk of individual spins within the spin-polarized ensemble leads to the random precession of spins around momentum-dependent internal magnetic field $\mathbf{B}_{so}$ and, thus, opening a pathway for spin relaxation. By tuning of the sample spin-orbit interaction by changing the sample parameters one can tailor the electron spin coherence. See, for example, the calculations in Ref.[5] using the sample parameters like QW width, symmetry, and electron density. However, for the carriers confined within the density of donor states, having zero average wave vectors, the Dyakonov-Perel spin relaxation mechanism does not work. Instead, the randomized magnetic field induced by the SOC leads to the DP like spin relaxation through the spin hopping process between the donor sites or via exchange interaction between the spin states localized on the adjacent donors.[15]

A number of experimental investigations on the temperature influence of spin dynamics has been carried out in semiconductor QWs,[16,17] however, to our knowledge, none of the report using multilayer structure has, to date, been appeared in the literature. For the present investigation, we selected the triple quantum well (TQW) because such multilayer structures lead to the discoveries of remarkable phenomena such as the drift of long current-induced spin coherence[18,19] and collapse of the quantum Hall interlayer tunneling gapes.[20] Additionally, such

[a)]Corresponding author.
Electronic address: saeedullah.phy@gmail.com

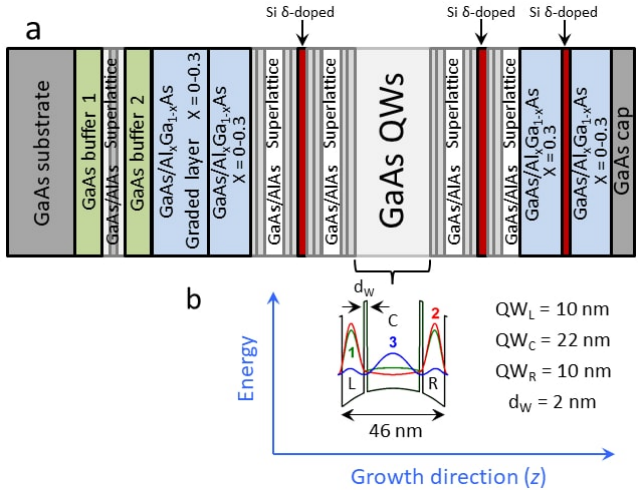

FIG. 1. (a) Schematic layer structure of the triple quantum well grown by MBE along $\hat{z}\|[001]$. (b) TQW band structure and charge density for the three occupied subbands with subbands separation $\Delta_{12} = 1.0$ meV, $\Delta_{23} = 2.4$ meV, and $\Delta_{13} = 3.4$ meV.

structures also offer possibilities for the generation of spin devices, for example in the production of spin filters.[21] Despite the fact that in systems with two or more occupied subbands the intra- and intersubband SOC may also suppress the spin coherence, the studied structure shows the robustness of spin polarization against temperature. Remarkably, it results $T_2^*$ in the nanosecond range for the studied structure with individual subband density beyond the metal-insulator transition.

## II. MATERIALS AND EXPERIMENT

The experiments were carried out on a high mobility $n$-GaAs/AlGaAs TQW grown by molecular beam epitaxy (MBE) on a (001)-oriented GaAs substrate. The layered structure of the sample is shown schematically in Fig. 1(a). The sample was remotely $\delta$-doped where three doping layers were deposited into the barrier materials of the quantum wells. The doping layers close to the left and right of the QWs provide carriers for the two-dimensional electron gas (2DEG) while the third doping was carried out to saturate the dangling bonds on the structure. The electrons from the doping layers were being collected into QWs forming a dense 2DEG with total electron sheet density of $n_s = 7 \times 10^{11}$ cm$^{-2}$. The sample growth condition was optimized to yield a 22-nm-thick GaAs central well and two 10-nm-thick lateral wells sandwiched between AlGaAs layers. The side wells are separated from the central well by 2-nm-thick Al$_{0.3}$Ga$_{0.7}$As barriers. The optimization was found necessary because the electron density mostly concentrates in the side wells as result of electron repulsion and confinement. The calculated band structure and subband charge density are illustrated in Fig. 1(b), where three subbands, with subband separation of $\Delta_{12} = 1.0$ meV, $\Delta_{13} = 3.4$ meV and $\Delta_{23} = 2.4$ meV, are formed as a result of interlayer coupling.[22]

The time-resolved Kerr rotation (TRKR) and resonant spin amplification (RSA) techniques were applied to demonstrate the long-lived spin coherence and its robustness against temperature. Both pump and probe pulses were delivered by Ti-sapphire laser with a pulse duration of 100 fs, operating at a frequency of $f_{rep} = 76$ MHz. The polarization of pump pulse was modulated by a photo-elastic modulator (PEM) operated at 50 kHz for lock-in detection. The circularly polarized pump pulse was focused onto a spot of approximately 50 $\mu$m on the sample. For all the experiments, except power dependence, we used the pump power of 1 mW (corresponding to excitation density of 50 W/cm$^2$) which give rise to the photogenerated carrier density of 2.0 $\times 10^{11}$ cm$^{-2}$. Varying the time delay $\Delta t$ between pump and probe pulses the rotation of linearly polarized probe upon reflection from the sample surface was recorded using a balanced bridge and double lock-in detection technique.

## III. RESULTS AND DISCUSSIONS

### A. Magnetic field dependence of spin dynamics

For the selection of right excitation energy, we first measured the TRKR dependence on the laser wavelength. Fig. 2(a) shows a series of TRKR traces recorded for different excitation energy with a wavelength ranging from 811 nm to 823 nm, under an applied magnetic field of $B = 1$ T. Obviously, the damping of spin beats and hence $T_2^*$ varies with the laser detuning when occupying the conduction and donor band states. The magneto-photoluminescence spectra of the studied structure can be found in Ref.[20], where two distinct regions were pointed out. The region at low wavelengths was correlated to the direct recombination between the states confined in the conduction and valence bands, where the one at high wavelengths was associated to an exciton bound to a neutral donor (DX center). For the conduction band states (low wavelengths), $T_2^*$ is smaller, whereas for higher wavelengths, at which the donor states are pumped, $T_2^*$ is much longer and persist between successive pulses as evidenced by non-vanishing signal at $\Delta t < 0$. Our data are in agreement with a similar study reported in $n$-type bulk GaAs doped beyond MIT, where, a strong variation of $T_2^*$, about three order of magnitude, as a function of excitation energy was found when occupying donor and conduction band states.[23] For the robustness of spin polarization, we chose the DX transition energies as these energies yield long-lasting spin signals. Additionally, for $\lambda = 823$ nm one can clearly see that the KR amplitude increases with time delay. In spatially-resolved Kerr rotation, such a behavior was attributed to the out-diffusion of photo-generated spins from the region of the laser spot.[24] However, in the present case, such a contribution was restricted by increasing the size of the spot to 50 $\mu$m. In our experiment such an effect originates due to the anti-phase contribution from the previous pulse.

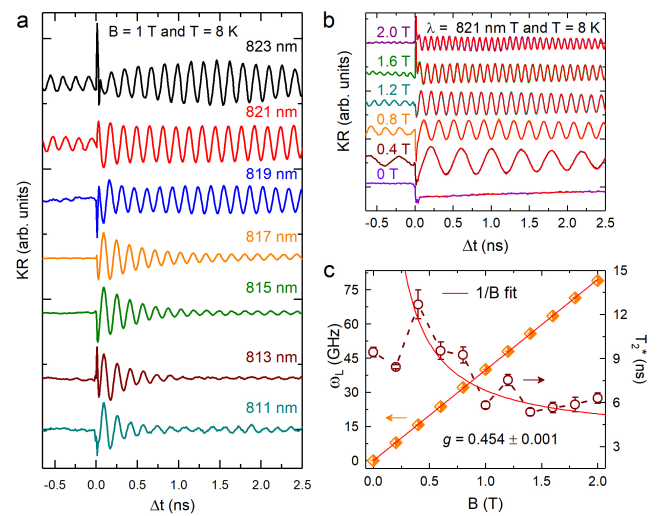

FIG. 2. (a) KR as a function of pump-probe delay measured for different wavelengths. (b) Fit to the Kerr rotation of optically induced spin polarization for various magnetic fields. (c) $T_2^*$ and $\omega_L$ as a function of applied magnetic field where solid red lines are fit to the data.

Fig. 2(b) shows the time evolution of Kerr rotation recorded with and without external magnetic field up to 2 T applied along x, i.e., in the Voigt geometry. TRKR traces (measured at $T = 8$ K and $\lambda = 821$ nm) show periodic oscillation in the external magnetic field, denoting the existence of spin signals. These oscillation results from the spin precession around the applied in-plane magnetic field with a beating frequency ($\omega_L$). Increasing the magnitude of applied magnetic field speed up the precessional frequency as evidenced from the TRKR traces. Furthermore, the decay of spin beats amplitude is very slow lasting more than the period of laser pulses ($t_{rep} = 13.2$ ns).

The measured TRKR signals are well described by the following function:

$$\Theta_K = A exp\left(\frac{-\Delta t}{T_2^*}\right) cos(\omega_L \Delta t + \phi) + y_0 \qquad (1)$$

where $A$ is the initial spin polarization amplitude, $\Delta t$ is the time delay between the pump and probe pulses, $T_2^*$ is the ensemble dephasing time, $\phi$ is the initial phase, $y_0$ is the Kerr signal offset, and $\omega_L = |g|\mu_B B/\hbar$ is the Larmor precession frequency with electron g-factor $|g|$, Bohr magneton $\mu_B$, magnetic field $B$ and reduced Planck's constant $\hbar$. The experimental curves were fitted to an exponential decay function for $B = 0$ and exponentially decaying cosine function (Eq: 1) for $B \neq 0$ as shown by red curves plotted on the top of experimental data. $\omega_L$ (for $B \neq 0$) and $T_2^*$ retrieved from fit are displayed in Fig. 2(c). As expected, $\omega_L$ varies linearly with applied magnetic field which is typical for the electrons[5,18], however, for holes, non-linearities can occur due to band mixing as reported for GaAs/In$_x$Ga$_{1-x}$As QWs.[25] The slope (solid red line) yields a $g$-factor (absolute value) of $g = 0.454 \pm 0.001$, where, its comparison with the bulk $|g|$ value further supports that the observed signals correspond to electron carriers.

The spin dephasing time varies with growing magnetic field see, for example, Fig. 2(c). $T_2^*$ first increases to a maximum value of $\sim 12.7$ ns at $B = 0.4$ T and then decreases with further increase of magnetic field due to the spread in ensemble $g$-factor.[5,26] The observed increase may be caused by the cyclotron motion acting as a momentum scattering, which in agreement with the Dyakonov-Perel mechanism, lead to a less efficient spin relaxation.[14] The reduction of $T_2^*$ with magnetic field follows $1/B$ dependence, where, the size of inhomogeneity can be inferred from the linear dependence of relaxation rate on the applied magnetic field, $1/T_2^* = \Delta g \mu_B B/2\hbar$.[26] We evaluated $\Delta g = 0.0005$, which is only 0.11 % of the observed $g$-factor suggesting that the spread of ensemble $g$-factor is not the only mechanism responsible for the spin relaxation. However, in QDs, $\Delta g$ can be quite sizable and can result in an efficient dephasing.

### B. Temperature influence on spin dynamics while tuning laser energy to donor states

A representative selection of TRKR traces measured for various temperatures in the range from 5 K to 250 K are shown in Fig. 3(a). For clarity of presentation, the TRKR traces are vertically shifted, and the curves at higher temperature are upscaled by multiplying with indicated numbers labeled inside the panel. To highlight the trends at negative $\Delta t$, the curves are normalized to the time origin ($\Delta t = 0$). One can clearly see significant changes in the carriers spin precession with rising temperature as highlighted by vertical dashed lines. The precession frequency is slowing down with increasing temperature, and the decay of spin beats amplitude is thermally stimulated. Additionally, the curves at low temperature (5-50 K) look phase shifted by $\Pi$ with respect to 140 and 250 K curves due to the generation of initial spin polarization in the opposite directions. This phase shift may be possibly due to the contribution of hole spin polarization to that of the electrons in the initial few picoseconds. Such a shift can also be seen in the magnetic field dependence where the hole contribution is evidenced by the shift of the center of gravity of the carrier spin precession. Also, at low temperatures in the range from 5 K to 35 K, the electron spin beating at positive delays are accompanied by spin beating even at negative delays due to the long-lived spin coherence persisting between successive pulses. In such cases, the spin dephasing time in excess of $t_{rep}$ can be retrieved by

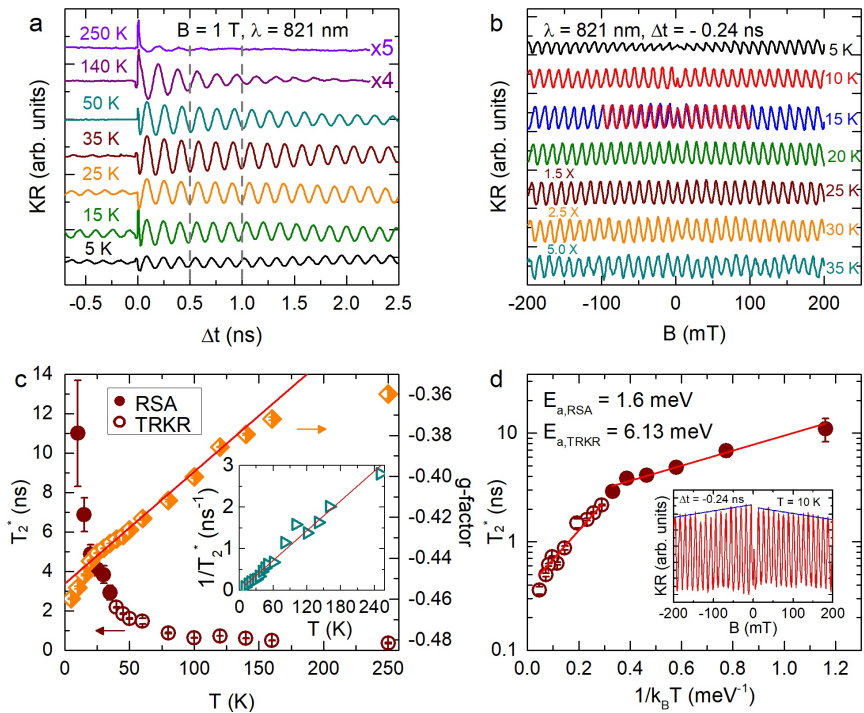

FIG. 3. Temperature influence on spin dynamics (a) Spin precession measured at various temperature in the range from 5 K to 250 K. (b) RSA measured at $\Delta t = -0.24$ ns for different temperatures. (c) The electron $g$-factor and $T_2^*$ extracted from the RSA (solid circles) and TRKR (open circles) as a function of temperature. The solid red line is linear fit to the data. The size of error bars shows the uncertainty in the measured values. The inset shows the temperature dependence of spin relaxation rate with linear interpolation (red line). (d) $T_2^*$ as a function of $1/k_BT$ fitted to Arrhenius-like function (solid lines). The measurement parameters are listed inside corresponding panels. Inset Fig. 3(d) highlights the decrease of the amplitude of finite field RSA peaks with increasing magnetic field.

using the RSA technique.[9]

Fig. 3(b) displays the RSA pattern recorded at $\Delta t = -0.24$ ns while sweeping the magnetic field over a range of -200 mT-200 mT. We observed sharp resonance peaks with spacing $\Delta B$ corresponding to the spin precession frequencies which are commensurable with the pulse repetition period obeying the periodic condition $\Delta B = h/g\mu_B t_{rep}$.[9] $T_2^*$ can be directly evaluated from the width of those peaks using Lorentzian model:

$$\Theta_K = A/\left[(\omega_L T_2^*)^2 + 1\right] \qquad (2)$$

where the half-width $B_{1/2}$ of RSA peaks point to the spin dephasing time $T_2^* = \hbar/g\mu_B B_{1/2}$. From the RSA spectrum, the following significant features can be directly extracted. First, in the temperature range from 5 K to 20 K the RSA peaks centered at $B = 0$ are smaller in amplitude than the peaks at $B \neq 0$. The depression of these zeroth-field resonant peaks are due to the spin relaxation anisotropy[27,28] caused by internal magnetic field. The direction and magnitude of this internal field can be obtained by fitting the data to the model formulated in Ref.[29], for example, as shown in Fig. 3(b) by red curves (in a selective range from -100 to 100 mT) for $T = 15$ K. The fitting yields the magnitude of internal magnetic field $B_\perp = 0.0017$ mT which causes spin relaxation in the material. More detailed analysis of anisotropic spin relaxation, internal magnetic field and it's influence on experimental parameters are published in Ref.[30].

Second, the data support a transition from anisotropic to isotropic spin relaxation with growing temperature i.e. with the rise of temperature the amplitude of the peak centered at $B = 0$ increases and become equal to that of finite field peaks at a higher temperature. Third, the amplitude of the finite field RSA peaks decrease with increasing magnetic field due to the ensemble spread of electron $g$-factor as noted above. However, plotting a series of RSA curves may be hiding this trend. See for example the RSA curve (inset Fig. 3(d)) recorded at $T = 10$ K where the trend is clearly visible. Fourth, at $T = 5$ K the resonant peaks have larger amplitude at higher magnetic field which is a well-known indication of the long hole spin coherence time involved in the generation of spin coherence time.[31]

$T_2^*$ received from the Lorentzian fit to the RSA peaks are depicted in Fig. 3(c) by closed circles together with the data points extracted from TRKR (open circles). The TRKR signal recorded at $T = 250$ K showed a biphasic spin dynamics and was fitted to Eq. 1 plus a non-oscillatory exponential decay to account the fast decay over first few picoseconds. The fitting yields decay times with $T_{2_1}^* = 8.5$ ps (corresponding to the hole spin dynamics) and a relatively long $T_{2_2}^* = 0.357$ ns (related to the electron spin dynamics). In the studied temperature range, $5$ K $< T <$ $250$ K, the electron $g$-factor increases from -0.46 to -0.36 (see Fig. 3(c)). For the temperature up to 160 K, a good linear dependence on temperature, $g(T) = -0.452 + 5.37 \times 10^{-4}T$, was observed. Our findings are in good agreement with a similar investigation reported on bulk GaAs.[32,33] In Ref.[32] the experimentally observed $g$-factor was approximated by $g(T) = -0.44 + 5.0 \times 10^{-4}T$ for a temperature ranged from liquid helium temperature up to room temperature.

We observed a strong $T_2^*$ reduction with temperature, decreasing down to 0.36 ns at $T = 250$ K. $T_2^*$ versus temperature shows a double linear dependence which is even more pronounced when dephasing time is plotted as a function of reciprocal thermal energy ($1/k_B T$). The observed dependence was attributed to the different spin relaxation mechanisms active at respective temperature ranges. At low temperatures, the spin relaxation was caused by exchange interaction with other localized donor states. However, at high temperatures, the spin relaxation was governed by carriers hopping process between the nearby donor sites.[12] In the presence of randomized spin-orbit field, both the exchange interaction between the spin states localized on the adjacent donors and the spin hopping process may lead to the change of electron spin states. And therefore, in principle, they can result in the spin relaxation. These processes can lead to a linear increase of the spin dephasing rate ($1/T_2^*$), in a similar way as for the classical DP mechanism, with temperature. Such a linear increase of relaxation rate, received from the TRKR and RSA, is shown in the inset Fig. 3(c). Fig. 3(d) shows the relaxation times as a function of reciprocal thermal energy fitted to the Arrhenius law: $A exp(E_a/k_B T)$, where A is the Arrhenius free-exponential factor, $E_a$ is the activation energy, and $k_B$ is the Boltzmann constant. Fit to the data yield the activation energies, labeled inside the corresponding panel, which are attributed to the hopping process between the donor sites.

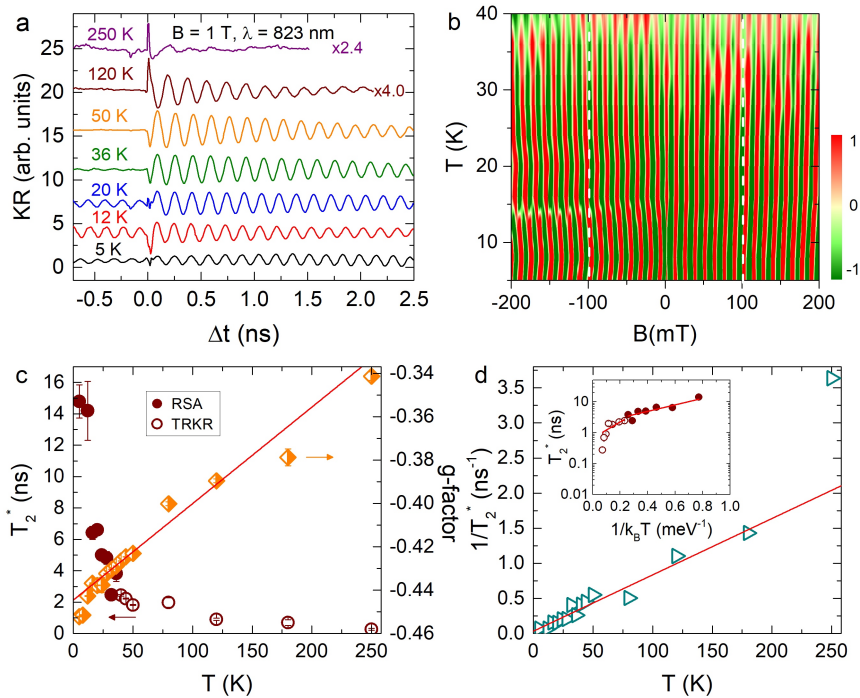

FIG. 4. Temperature influence on spin dynamics (a) KR vs $\Delta t$ recorded at different temperature. (b) Temperature dependence of resonant spin amplification. (c) The $g$-factor and $T_2^*$ as a function of sample temperature where $T_2^*$ retrieved from the RSA signal are shown by solid circles while open circles depict the one obtained from the TRKR. (d) Temperature dependence of $1/T_2^*$ with a linear fit to the data (red line). Inset shows $T_2^*$ as a function of reciprocal thermal energy fitted to Arrhenius-like function.

Concerning the subband dependence of spin dynamics, the laser energy was changed by about 3 meV ($\simeq \Delta_{13}$) by increasing the pump-probe wavelength from 821 to 823 nm. Figure 4(a) shows a set of TRKR traces measured at different temperatures while tuning the laser wavelength to 823 nm and keeping the experimental conditions the same as were used in the previous section for maximum KR signal. One can clearly see that the precession frequency is getting smaller with growing temperature which directly affects the electron $g$-factor through the relationship described in Sec. III A. As exhibited, all the data except the ones at elevated temperatures show a single exponential decay. However, for the present study, only the spin relaxation of electrons are concerned, while the short relaxation times (such as hole spin polarization) are disregarded. The pronounced oscillation at negative time delays, of the amplitude comparable to the one at positive delays, observed at low temperatures suggest that $T_2^* \geq t_{rep}$.

In analogy to the previous discussion, we used the RSA technique[9] to extract $T_2^*$, which takes into account the constructive interference of the coherent spin oscillations from successive pulses. Such an RSA pattern measured at $\Delta t = -0.24$ ns is shown in Fig. 3(b). The superposition of spins that were created by the pulse train 13.2 ns before the arrival of the next pulse causes a series of sharp resonance peaks as revealed by a striped pattern. The rise of temperature accelerate the decay of spin polarization due to heating and the RSA peaks disappear into noise (white shades) at higher temperature see for example the peaks at $T \geq 35$ K. Additionally, the variation of $g$-factor is clearly evidenced by the change of the spacing, $\Delta B$, between RSA peaks. That is with growing temperature the outer peaks are shifting toward higher magnetic fields as marked by white dashed lines at $B = \pm 100$ mT.

The electron $g$-factor, received from the TRKR oscillation, increase linearly with a slope of $4.87 \times 10^{-4} K^{-1}$. Again, the present findings are in agreement with the literature results.[32,33] $T_2^*$ extracted from the RSA (solid circles) and TRKR (open circles), plotted in Fig. 4(c), shows a strong reduction with temperature. Fig. 4(d) shows the temperature dependence of $1/T_2^*$ following a linear increase, up to 190 K, with a slope of 0.008 ns$^{-1}$K$^{-1}$. The trends of shortening $T_2^*$ above 190 K, deviates from the linear behavior, suggest that the higher temperature causes the heating effect which leads to low spin polarization as commented above. The inset shows $T_2^*$ as a function of $1/k_BT$, fitted to Arrhenius-like function yielding activation energies of 2.43 meV and 6.01 meV for the RSA and TRKR respectively. The difference in $T_2^*$, while changing the laser energy by about 3 meV ($\simeq \Delta_{13}$), may be associated to the relative different charge density distribution of electrons in the first and third subbands.

### C. Dependence of spin dynamics on optical power

In this section, we report on the excitation power influence on the spin dynamics measured by using time-resolved Kerr rotation. Fig. 5(a) shows the pump-probe delay scans of the KR signal measured, at $B = 1$ T applied normal to the initial spin polarization, for different excitation powers in the range from 1 mW to 7 mW (corresponding to 50-350 W/cm$^2$). At a low pump power of 1 mW the density of photogenerated carriers is comparable to the 2DEG density, however, at high power, the photogenerated density exceeds the density of 2DEG by an order of magnitude. The striking feature of the KR traces is the appearance of a long-lived spin beating as can be seen at negative time delay. The electron $g$-factor evaluated from the fit of experimental data are shown in Fig. 5(b). As expected we didn't see any influence of the optical pump power on the $g$-factor which is directly reflected from the constant spin beats frequency marked by dotted lines in panel (a).

The resulted values of $T_2^*$, plotted in Fig. 5(c), remain constant in the low power range. However, further increase of excitation power results in the decrease of $T_2^*$. For single QW structure, the reduction of $T_2^*$ at high pump density was assigned to the heating effect induced by optical excitation.[34] In our structure, we attribute this decrease to an increased efficiency of Bir-Aronov-Pikus (BAP) mechanism induced by the high density of photogenerated carriers. As a key factor for practical spintronics, we noticed that $T_2^*$ still remains in the nanosecond range when the excitation power is raised by almost one order of magnitude. The observed long-lasting $T_2^*$ results from the simultaneous suppression of BAP and spin-orbit relaxation mechanisms governed by spin hopping and exchange between adjacent donor sites.

## IV. CONCLUSIONS

In summary, we have studied the magnetic field, sample temperature, and optical pump power dependence of spin dynamics in a GaAs/AlGaAs triple quantum wells by using pump-probe Kerr rotation. It has been found that the spin polarization in our sample is robust against temperature and was apparent up to $T = 250$ K. The increase of excitation energy about 3 meV ($\simeq \Delta_{13}$), by varying the laser wavelength from 823 nm to 821 nm, causes a $T_2^*$ reduction of 25 % at $T = 250$ K. The spin-orbit relaxation powered by spin hopping process or exchange interaction between the states localized on nearby donors lead to a linear increase of dephasing rate on temperature. Additionally, the electron $g$-factor was also noticed to vary linearly with temperature. This behavior is in agreement with the data reported on bulk GaAs.[32,33] The observation of long-lived spin coherence persisting up to high temperature, and the spin relaxation anisotropy, adding the attractiveness of multilayer structure for practical spintronics.

### ACKNOWLEDGMENTS

F.G.G.H. acknowledges financial support from Grant No. 2009/15007-5, 2013/03450-7, 2014/25981-7 and 2015/16191-5 of the São Paulo Research Foundation (FAPESP). S.U acknowledges TWAS/CNPq for financial support.

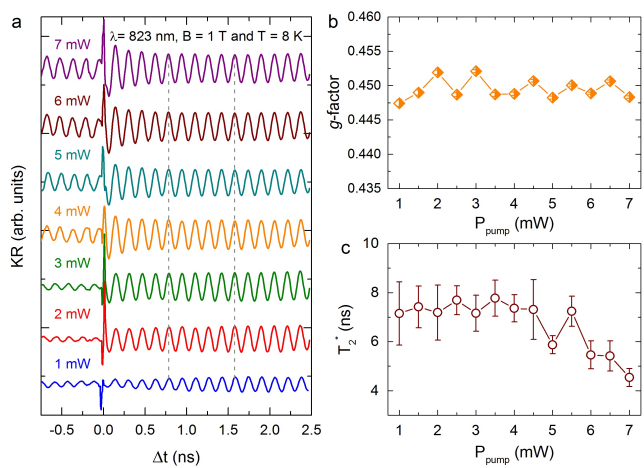

FIG. 5. Pump power influence on the spin dynamics: (a) TRKR signals as a function of excitation power. The evaluated (b) $g$-factor and (c) $T_2^*$ as a function of pump power. The measurement parameters are listed in panel (a).